# Synthesis and electronic structure characterization of diamane

Feng Ke[1,2,3#], Lingkong Zhang[1#], Yabin Chen[4#], Ketao Yin[5], Chenxu Wang[2], Wanquan

Zhu[6], Hailun Wang[1], Yu Lin[3], Zhenxian Liu[7], John S. Tse[5], Guilin Wu[6], Rodney C.

Ewing[2], Wendy L. Mao[2,3], Junqiao Wu[4], Ho-Kwang Mao[1] & Bin Chen[1]*

*1Center for High Pressure Science and Technology Advanced Research, Shanghai

201203, China*

*2Department of Geological Sciences, Stanford University, Stanford, CA 94305, USA*

*3Stanford Institute for Materials and Energy Sciences, SLAC National Accelerator

Laboratory, Menlo Park, CA 94025, USA*

*4Department of Materials Science and Engineering, University of California,

Berkeley, California 94720, USA*

*5Department of Physics and Engineering Physics, University of Saskatchewan,

Saskatoon, Saskatchewan, S7N 5E2, Canada*

*6Joint International Laboratory for Light Alloys, College of Materials Science and

Engineering, Chongqing University, Chongqing, 400045, China*

*7Institute of Materials Science, Department of Civil and Environmental Engineering,

The George Washington University, Washington, DC 20052, USA*

[#] F.K., L.Z., and Y.C. contributed equally to this work

*To whom correspondence should be addressed. E-mail: chenbin@hpstar.ac.cn

## Abstract

Atomically thin graphite, known as graphene, has been a marvel in material science because of its exceptional properties, novel physics and promising applications. Atomically thin diamond, called diamane, has also attracted considerable scientific interest due to its potential physical and mechanical properties. However, until now there has been no reports of successful synthesis of a free-standing pristine diamane film. Here, we report the synthesis and electronic structure characterization of diamane. Electrical measurements, x-ray diffraction and theoretical simulations reveal that trilayer and thicker graphene transform to hexagonal diamane (h-diamane) when compressed to above 20 GPa, which can be preserved down to few GPa. Raman studies indicate that the sample quenched from high pressure and high temperature also has a h-diamane structure, i.e., h-diamane is recovered back to ambient conditions. Optical absorption and band structure calculations reveal an indirect energy gap of $2.8\pm0.3$ eV in the diamane film. Compared to gapless graphene, diamane with sizable bandgap may open up new applications of carbon semiconductors.

Although diamond and graphite are both allotropes of carbon, their very different structures and bonding lead to dramatically different physical, chemical and mechanical properties. Diamond is the hardest bulk material, the best thermal conductor, is chemically inert and optically transparent compared to graphite which is soft, opaque semimetal. Atomically thin graphite, i.e., graphene, has been found to have many exceptional physical properties, such as high carrier mobility[1], half-integer quantum Hall effect[2-4], unconventional superconductivity[5]. Atomically thin layer of diamond, if it could be synthesized, would be predicted to have dramatically different properties from graphene.

Previous theoretical and experimental studies suggest that atomically thin diamond films do not exist in free or pristine state due to the lack of thermally stable two-dimensional structure but are achievable if the surfaces are thermodynamically equilibrated with specific chemical groups such as hydroxyl or hydrogen[6-13]. Surface hydrogenation or fluorination for synthesizing such diamond films, called diamondene[6], diamondol[7], diamane[8,10], or diamene[13] have been attempted, and use of various substrates such as Co, Ni, Cu, SiC have also been introduced in these attempts[13,14]. The substrates have been shown to regulate the physical properties of graphene significantly[6,15-17]. It was recently reported that diamond-like carbon was observed when micro-indenting 2-layer graphene with the surface carbon atoms interacting strongly with the Si-face of SiC substrate[13]. The conversion was reversible and not observed in thicker (3- and 5-layer) graphene. Furthermore, all these attempts changed the nature of the materials.

Despite significant effort being invested in the synthesis of atomically thin diamane films, there have not been a report of successful synthesis of a free-standing pristine diamane film. Here we used an alternate approach, the diamondization of mechanically exfoliated few-layer graphene via compression, a clean method which does not introduce any chemical impurities, to synthesize the long sought diamane films. Diamane has several polytypes due to different stacking sequence of carbon layers. For simplicity, we compactly classified h-diamane with the number of carbon layers[10], i.e., 3, 4, 6-layer (3L, 4L, 6L) h-diamane are transformed from trilayer, tetralayer and hexalayer graphene, respectively. We demonstrate that (n ≥3)-layer h-diamane are successfully synthesized by compression of trilayer and thicker graphene, which could be preserved to few GPa under decompression. In additional our finding reveal that h-diamane synthesized at high pressure and high temperature could be recovered back to ambient conditions.

Electrical measurements are a sensitive probe to study the pressure-induced graphite to diamond transition since the $sp^2$-$sp^3$ rehybridization between carbon atoms is accompanied by the opening of an energy gap and a dramatic increase in resistance[18,19]. Using our recently developed photolithography-based micro-wiring technique to prepare electrodes on diamond surface for atomically thin samples (Fig. 1a)[20], we studied the pressure-induced diamondization process of few-layer graphene by measuring the sheet resistance.

High-pressure resistance measurements were conducted on graphene ranging in layer thickness from multilayer (graphite, 1 μm in thickness) to bilayer at room temperature. All the samples except bilayer show obvious transition with dramatical increases of resistance (Fig. 1b). The transition pressure is significantly dependent on the layer number, i.e., the transition occurs at a higher pressure in thinner sample. Specifically, the sheet resistance of multilayer graphene varies smoothly until the onset of the transition just above 15.1 GPa (inset of Fig. 1b), followed by a substantial resistance increase of more than five orders of magnitude upon further compression to 55.0 GPa, beyond which the value exceeds the measurable range of our instruments, comparable with previous electrical results caused by the graphite to h-diamond transition under compression[18,19]. The transition pressure is comparable to those observed in other measurements[21-28]. In the case of few-layer graphene samples, the transition occurs at about 19.6, 21.3, 27.1 and 33.0 GPa in 12-layer, hexalayer, tetralayer and trilayer graphene (Fig. 1b), respectively, suggesting the formation of 12L, 6L, 4L and 3L h-diamane under compression. Upon decompression, the high-pressure phase can be quenched to few GPa, but goes back to the initial state after releasing to ambient pressure. For bilayer graphene, the sheet resistance value remains nearly constant with compression up to 60.0 GPa (Fig. 1b), the highest pressure studied in our measurements, suggesting that if it were to occur a higher pressure would be needed to drive the diamondization process.

The temperature dependence of sheet resistance of few-layer graphene samples confirm the layer dependence of diamondization transition (Extended Data Fig. 2). For all the graphene samples from multilayer to trilayer, their resistance shows weak positive temperature dependence ($dR/dT\approx0$) before the phase transition, indicating a semimetallic character. At higher pressures, such as trilayer graphene above 35.2 GPa, the temperature dependence of resistances becomes negative (d$R$/d$T$<0) and increasingly steep in further compression, signaling the semiconducting behavior. In bilayer graphene, however, the $R$-$T$ curves always show weak temperature dependence up to 60.0 GPa, behaving as a semimetal.

To track the diamondization process directly, we conducted high-pressure x-ray diffraction (XRD) measurements on mixed few-layer graphene powders (3-8 layers for single flake, Extended Data Fig. 3), given the weak XRD signal of single few-layer graphene flake. Two new diffraction peaks around (100) and (101) peaks appear above 28.5 GPa (Fig. 2a), and their intensities increase with pressure. Above 50.3 GPa, all the diffraction peaks of the starting graphene powders disappear, indicating that the structural transition has completed. The new phase can be indexed into h-diamond structure (Extended Data Fig. 4). Other predicted structures, such as M-[29], H-[30], R-[31], W-[32], and Z-carbon[33], cannot fit the observed XRD pattern. The XRD pattern of the high-pressure phase is similar to previously reported XRD results of bulk graphite quenched from high pressure and high temperature[34], in which nanocrystalline h-diamond was observed with small fraction of cubic diamond mixed. During decompression, the sample goes back to its initial state after quenching to ambient pressure.

We performed extensive structure searches through the CALYPSO methodology to rationalize the pressure-induced structure evolution of few-layer graphene. In our simulations, 3L h-diamane in (-2110) orientation with a sizable energy gap, as shown in Fig. 3a, is found to be energetically stable and has a lower total energy than trilayer graphene above 50 GPa (Fig. 3 and Extended Data Fig. 5). In this structure, parts of the surface carbons are $sp^2$-like coupling with bonding lengths of 1.354 Å, and the rest surface and inner carbons $sp^3$-like hybridization (Fig. 3b). The angle between $P_{123}$ (the plane of the 1, 2 and 3 carbon atoms, Fig. 3b) and $P_{234}$ is about 143°, much larger than that of $P_{567}$ and $P_{678}$ (128°), which makes the surface carbon layers relatively flat compared to inner carbon layers. All these features contribute to the stability of (-2110)-oriented h-diamane even without hydrogenation or fluorination on the surface. Similar transitions are also observed at lower onset pressures, e.g., 30 GPa for 4L diamane and 15 GPa for 6L diamane, respectively, showing good agreement with our electrical results. In the case of bilayer graphene, the 2L (-2110)-oriented h-diamane does not have a lower energy than bilayer graphene until above 160 GPa (Fig. 3d).

Previous experimental studies indicated that h-diamond could be synthesized at high pressure and high temperature and preserved to ambient conditions[18,21-24,34], which hints for a possible route to synthesize pressure quenchable h-diamane. We therefore heated the samples at high pressure before quenching to ambient conditions. Two quenched samples are obtained, one by compressing multilayer graphene (graphite, 1 μm in thickness) up to 22.0 GPa and then heating up to 1500 K, and the other by compressing mixed few-layer graphene powders (3-8 layers for single flake, Extended Data Fig. 3) up to 28.0 GPa and then heating up to 1700 K. Our Raman characterization of the quenched multilayer graphene sample show the signature vibration modes of h-diamond at about 1225 cm$^{-1}$, 1319 cm$^{-1}$ and 1552 cm$^{-1}$ (Fig. 2b), confirming the h-diamond structure[35]. The signature vibration modes of h-diamond are also observed in the quenched few-layer graphene powder samples with slight softening of the vibration modes, and weak signals of cubic diamond and residual few-layer graphene, indicating the synthesis of pressure quenchable h-diamane.

The inner carbons are $sp^3$-bonded in h-diamane while parts of the surface carbons are $sp^2$-bonded. Their electronic structure should deviate substantially from those of graphene and diamond. For examining the electronic structure, infrared absorption measurements (the photon energy range of 0.2~1.0 eV) were conducted on graphene samples with thickness ranging from multilayer to trilayer under compression. The absorbance of all the graphene samples, A = -log $T$ drop suddenly above the onset pressure and approach to zero with further compression (Extended Data Fig. 6), indicating that the bandgap is opened by at least 1.0 eV.

To further track the opening of the bandgap, we measured the Vis-UV (1.4~4.9 eV) absorption spectra of trilayer and thicker graphene under compression. For all the graphene samples, pronounced and asymmetric peaks at a photon energy of 4.6 eV are observed in the absorption spectra at low pressures (Fig. 4b), which are the excitonic resonance and the feature of interband transition in graphene[36]. These peaks are insensitive to pressure until the structural transition occurred at which point their intensities drop dramatically. The disappearance of the excitonic resonance peaks confirm the loss of the electronic structure of trilayer and thicker graphene.

Above the graphene to h-diamane transition pressure, such as multilayer above 16.2 GPa, hexalayer above 20.6 GPa and trilayer above 28.3 GPa, a sharp drop of absorbance is observed at the radiation photon energy of 2.0 eV (Fig. 4c), followed by further decrease with pressure, which indicates that the bandgap is larger than 2.0 eV. The opening of bandgap is also in agreement with the optical microscopy observations that trilayer and thicker graphene becomes increasingly transparent above the transition pressure (Extended Data Fig. 7).

Weak absorption edges are observed at 2.8±0.3 eV in the absorption spectra of (n ≥3)-layer diamane. It is hard to identify the direct or indirect bandgap because of the weak absorption edges. Alternatively, we adopted the band structure calculations to identify the direct/indirect gap. The calculations indicate that h-diamane transformed from few-layer graphene has an indirect energy gap of 3.0±0.2 eV, with the valance band maximum and conduction band minimum located at Γ and S high symmetry points (Extended Data Fig. 8), respectively. The bandgap values show better agreement with the optical absorption results.

The experimental results suggest that a large pressure range is needed to make few-layer graphene transformed to h-diamane completely, i.e., the ratio of graphene/h-diamane is pressure dependent. Such phase mixture can be indicated from the non-uniform transparency of the compressed multilayer graphene. The transparent area expands gradually with pressure above the onset pressure. The continuous increase of resistance and decrease of absorbance with pressure, rather than sharp transitions at the onset pressure, can also be the consequence of the gradual transition from graphene to h-diamane transition.

In conclusion, free-standing pristine (n≥3)-layer h-diamane has been synthesized by compressing mechanically exfoliated trilayer and thicker graphene (up to 1 μm in thickness) at ambient temperature and once synthesized can be preserved down to few GPa. The transition pressure is significantly dependent on the layer number, i.e., the transition occurs at a higher pressure in thinner sample. H-diamane synthesized at high pressure and high temperature conditions, such as 25.0 GPa and 1700 K for mixed 3-8 layers graphene powders, could be recovered back to ambient pressure. An energy gap of $2.8\pm0.3$ eV is observed in ($n\geq3$)-layer h-diamane. Band structure calculations reveal the indirect gap with the valence band maximum and conduction band minimum located at Γ and S high symmetry points, respectively. The electronic bandgap usually determines the electrical and optical properties of a semiconductor, and further governs the operation of semiconductor devices such as field-effect transistors, lasers, etc. H-diamane semiconductor, in contrast to the gapless graphene, would enable potential applications of carbon-based electronic devices. After graphene, carbon nanotubes, fullerenes and other carbon allotropies, the realization of a free-standing pristine diamane would be another exciting achievement in material science, which may trigger many novel applications of carbon, such as its potential application in quantum computation.

**Acknowledgements**

We thank Drs. Hongwei Sheng, Xiao-Jia Chen, and Lin Wang for their helpful discussions and technical support. The authors acknowledge the funding support of NSAF (Grant No: U1530402). The device fabrication part done in Berkeley was supported by the U.S. NSF grant No. DMR-1708448. Y.L., W.L.M., and F. K.'s contribution to the experiments, data analysis and manuscript revising were supported by the US Department of Energy, Office of Science, Basic Energy Sciences, Materials Sciences and Engineering Division (DE-AC02-76SF00515).

**Author contributions**

B.C. and F.K. conceived of the project and drafted the manuscript. F.K., L. Z. and Y.C. fabricated the graphene devices, performed all measurements and data analysis. C.W. and Z.W. collected the TEM results. K.Y. conducted the simulations. All authors discussed the results and revised the manuscript.

The authors declare no competing financial interests.

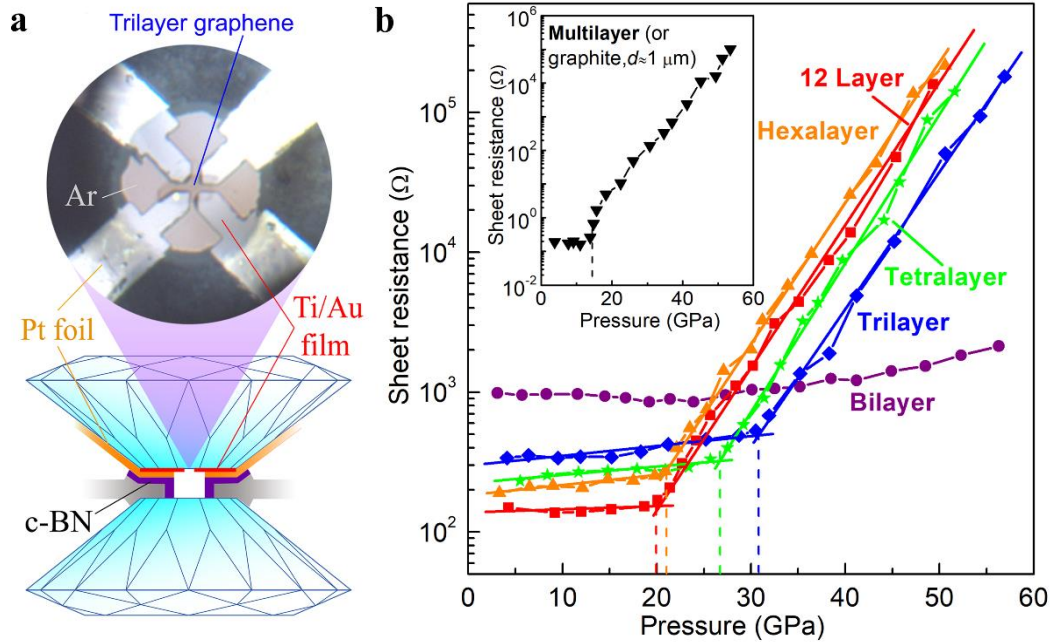

**Fig. 1. Sheet resistance of bilayer, trilayer, tetralayer, hexalayer and 12-layer as a function of pressure. a**, Photomicrograph of four-terminal nanodevice with trilayer graphene in diamond anvil cell from top view and schematic of designed microcircuit from cross-sectional view. For the four-probe electrode configuration, Ti/Au films are patterned onto the diamond culet and extended with the platinum foils. **b,** Resistance-pressure curves of bilayer, trilayer, tetralayer, hexalayer and 12-layer measured at room temperature. The solid lines are the guide for eyes. Inset: pressure dependence of resistance in multilayer graphene (graphite).

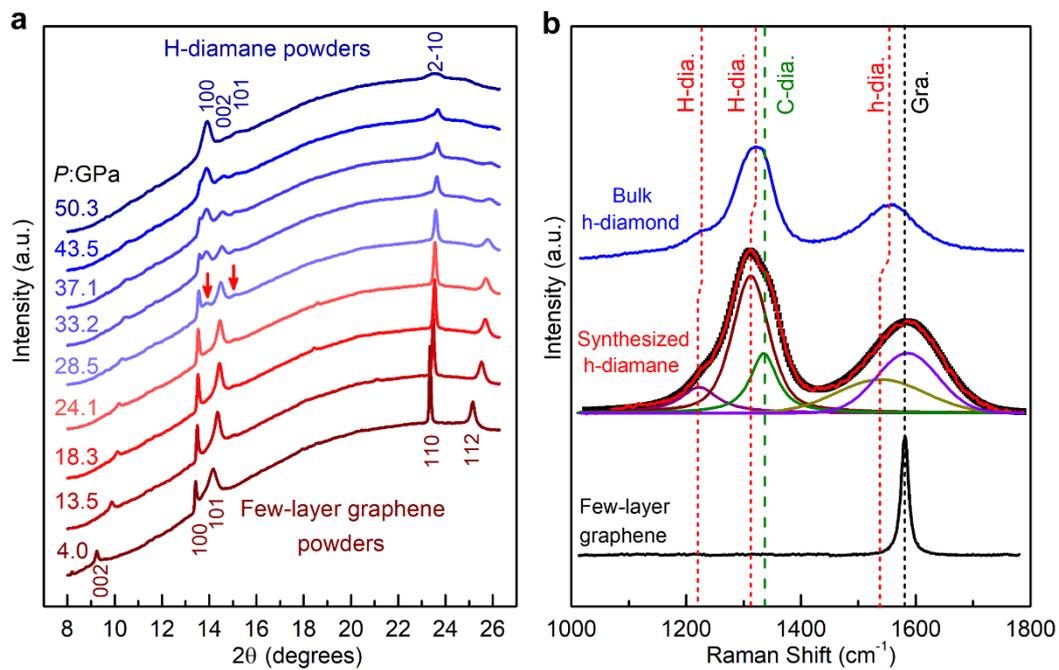

**Fig. 2. Experimental identification of the structure of diamane films. a**, XRD results of few-layer graphene powders under compression (λ=0.4959 Å). **b**, Raman spectra of quenched bulk h-diamond and few-layer diamane after high pressure and thermal treatment.

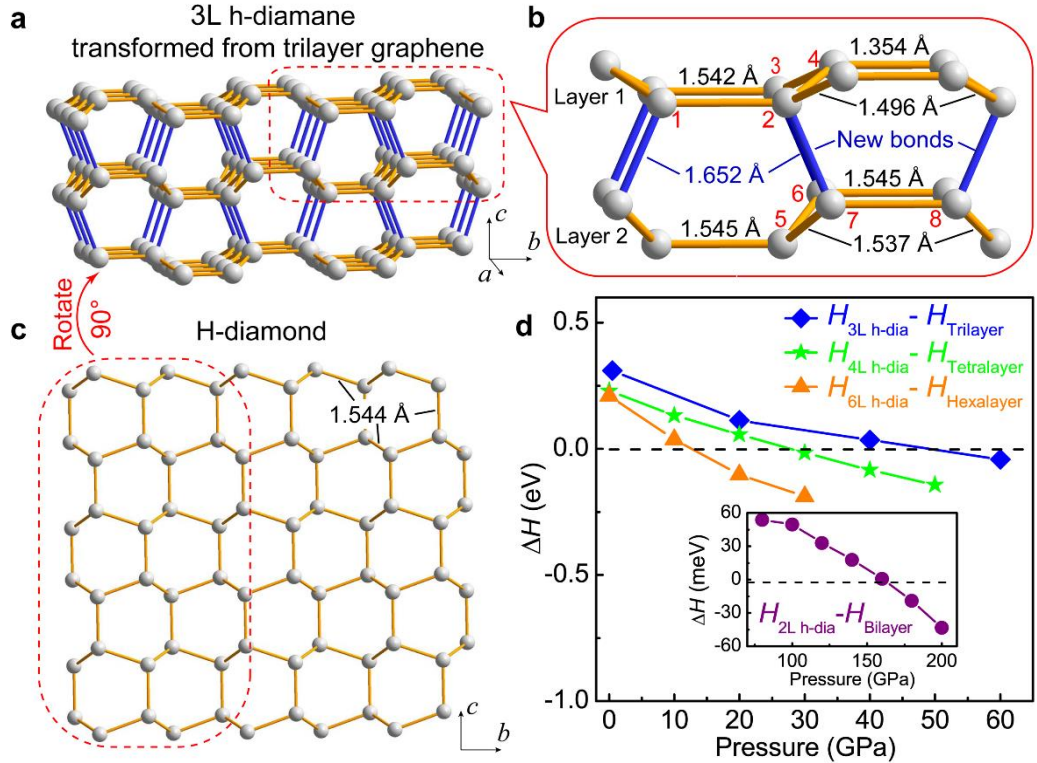

**Fig. 3. Structures and enthalpy calculations for h-diamane. a,** The structure of 3L
(-2110) oriented h-diamane transformed from trilayer graphene under compression. **b,**
The structural details of 3L (-2110) oriented h-diamane. **c,** The structure of h-diamond.
**d,** The enthalpies of 2L, 3L, 4L and 6L h-diamane (relative to bilayer, trilayer, tetralayer
and hexalayer graphene, respectively) as a function of pressure.

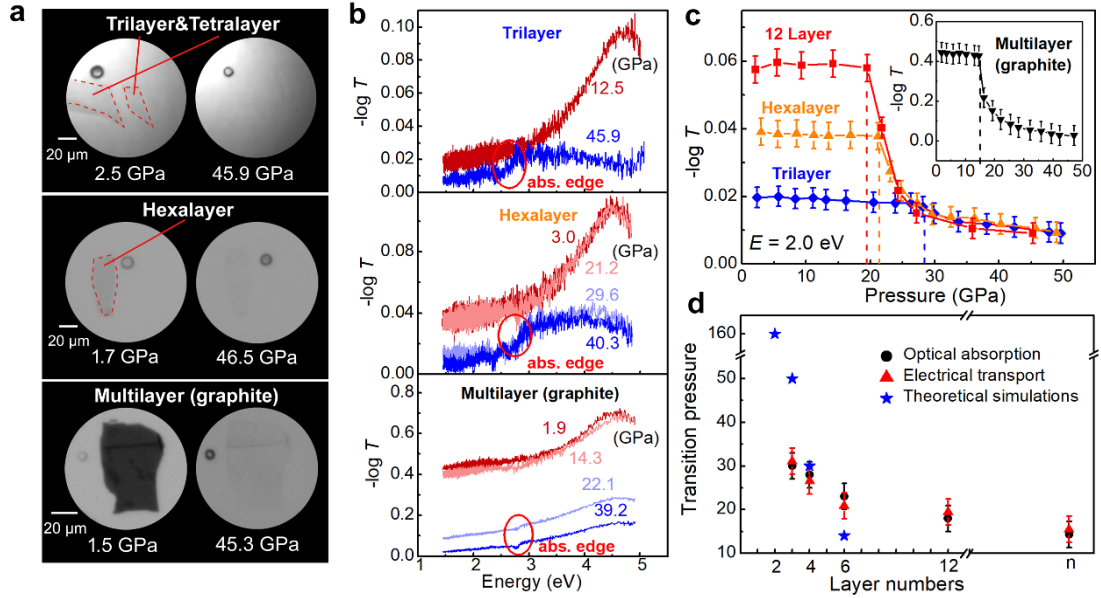

**Fig. 4. High pressure optical absorption measurements for few-layer graphene. a,** Photomicrographs of compressed trilayer, tetralayer, hexalayer and multilayer graphene (or graphite) samples in transmission mode with a white light source. **b,** Optical absorbance of trilayer, hexalayer and multilayer graphene as a function of pressure. The absorbance of the diamond and pressure medium is subtracted. **c,** Evolution of absorbance of trilayer, hexalayer and multilayer graphene (Inset) at a photon energy of 2.0 eV upon compression. **d,** The onset transition pressures of graphene to h-diamane as a function of graphene layers.

# Supplementary Information for

# Synthesis and electronic structure characterization of diamane

Feng Ke[1,2,3#], Lingkong Zhang[1#], Yabin Chen[4#], Ketao Yin[5], Chenxu Wang[2], Wanquan

Zhu[6], Hailun Wang[1], Yu Lin[3], Zhenxian Liu[7], John S. Tse[5], Guilin Wu[6], Rodney C.

Ewing[2], Wendy L. Mao[2,3], Junqiao Wu[4], Ho-Kwang Mao[1] & Bin Chen[1]*

*[1]Center for High Pressure Science and Technology Advanced Research, Shanghai

201203, China*

*[2]Department of Geological Sciences, Stanford University, Stanford, CA 94305, USA*

*[3]Stanford Institute for Materials and Energy Sciences, SLAC National Accelerator*

*Laboratory, Menlo Park, CA 94025, USA*

*[4]Department of Materials Science and Engineering, University of California,*

*Berkeley, California 94720, USA*

*[5]Department of Physics and Engineering Physics, University of Saskatchewan,*

*Saskatoon, Saskatchewan, S7N 5E2, Canada*

*[6]Joint International Laboratory for Light Alloys, College of Materials Science and*

*Engineering, Chongqing University, Chongqing, 400045, China*

*[7]Institute of Materials Science, Department of Civil and Environmental Engineering,*

*The George Washington University, Washington, DC 20052, USA*

[#] F.K., L.Z., and Y.C. contributed equally to this work

*To whom correspondence should be addressed. E-mail: chenbin@hpstar.ac.cn

## Methods

**Sample preparation and electrical measurements in diamond anvil cells.** Few-layer graphene flakes were mechanically exfoliated from bulk graphite on $Si/SiO_2$ surface. Optical microscopy, Raman spectroscopy and atomic force microscopy were utilized to identify the layer number of graphene samples. Few-layer graphene was then transferred onto diamond surface using a polydimethylsiloxane (PDMS) stamping technique[1]. During the transfer process, high-resolution microscopy and Raman spectra measurements were adopted to confirm that the graphene layer did not remain in the Si substrate or PDMS, *i.e.*, to guarantee that the graphene sample has been transferred onto the diamond surface without any damage. Four Ti/Au electrodes (150 nm in thickness) were then configured with photo-lithography and electron beam deposition techniques on diamond surface to make Ohmic contacts with the samples, as shown in Fig. 1a. Each Ti/Au film electrode was extended beyond the indentation area of gasket through hand-wiring platinum foil (~2 μm in thickness) electrodes to keep the stability of electrodes under pressure. The insulation of probing electrodes from the metallic rhenium gasket were done through a mixture of cubic boron nitride and epoxy pre-compressed into the indentation area of gasket. Daphne 7373 or argon pressure media were used to simulate a quasi-static pressure environment on samples. Low-temperature electrical measurements were conducted in PPMS with temperature range of 2-300 K.

**Raman characterization.** Raman spectra were collected with Renishaw InVia spectrometer. The identification of layer was done using a 532 nm laser as the incident light and 50x Leica optical microscope. The identification of h-diamond was done using the 325 nm laser as the incident light.

**Absorption measurements**. High-pressure Infrared absorption measurements were conducted using type-IIa diamonds at Beamline 1.4.3 of Advanced Light Source (ALS), Lawrence Berkeley National Laboratory (LBL) and Infrared Lab of National Synchrotron Light Source II (NSLS-II) at Brookhaven National Laboratory (BNL). Infrared spectra were collected on a Fourier transform infrared spectrometer coupled to a microscope with a mid-band MCT detector. Visible-ultraviolet absorption measurements were conducted on a customized visible-ultraviolet microscope system with the photon energy of 1.5-4.9 eV. The transmittance data of sample was obtained by two measurements, one being the transmission through the sample area ($T_s$) and the other through the empty area beside the sample ($T_0$), by which the background from diamond and pressure medium can be subtracted. For better comparison with electrical measurements, KBr, Daphne 7373 or argon was used as pressure media, respectively.

**Structure prediction calculations.** Structure prediction calculations were based on a global minimum search of the free energy surfaces obtained by ab initio total-energy calculations employing the CALYPSO methodology[2-4]. Structure searches were firstly performed on trilayer graphene at 0, 30 and 50 GPa. Usually, an empty space of ~10 Å will be set as the vacuum for simulation of 2-dimensinal structure, but which will be optimized to zero under pressure. In our high-pressure simulation, we used non-interacting monolayer graphene on both sides of trilayer graphene instead to achieve the high pressure. New energetically favorable structures were obtained from the structure search. After that, lattice optimizations were done on the obtained favorable structures with argon atoms surrounded to mimic hydrostatic condition. Band structure calculations were performed with the VASP codes using both the Perdew-Burke-Ernzerhof and heyd-Scuseria-Ernzerhof hybrid functional[5]. The projector-augmented wave method was used with $2s^22p^2$ (cutoff radius 1.5 a.u.) and $3s^23p^6$ (cutoff radius 1.9 a.u.) as valence electrons for C and Ar, respectively. The cutoff energy of 500 eV for the expansion of the wave function into plane waves and fine Monkhorst-Pack $k$ meshes were chosen to ensure that all the enthalpy calculations are well converged to better than 1 meV/atom.

**High-pressure XRD measurements**. High-pressure XRD experiments were performed at BL12.2.2 of the Advanced Light Source with the wavelength of 0.4959 Å . Few-layer graphene powdered samples, together with the methanol-ethanol (4:1) as transmitting medium, were loaded into the sample chamber. Daphne 7373 was also used as the medium for a convenient comparison of the XRD and electrical transport results.

**Laser heating experiments.** The few-layer graphene powders were heated by a double-sided laser system. The heating temperature was measured by fitting the black-body radiation curve.

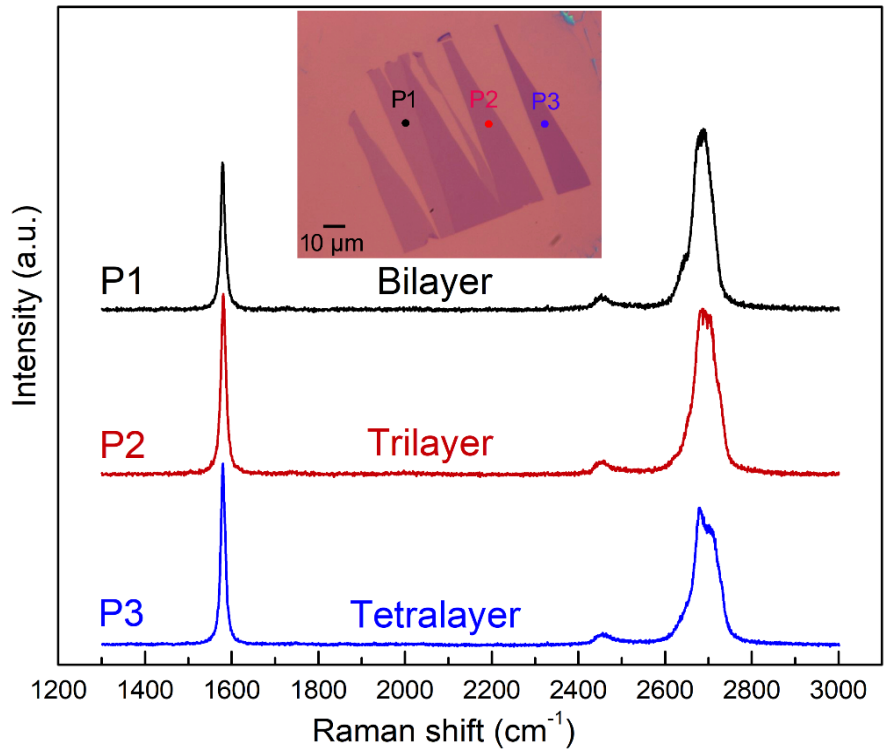

**Extended Data Figure 1 │ Identification of few-layer graphene by Raman and optical microscopy**. Raman spectra of few-layer graphene on Si/SiO₂ substrate. Few-layer graphene are identified by optical contrast in microscope and then confirmed by Raman spectra before transferring to diamonds for further measurements. The flake is of high quality without any D band around 1350 cm⁻¹.

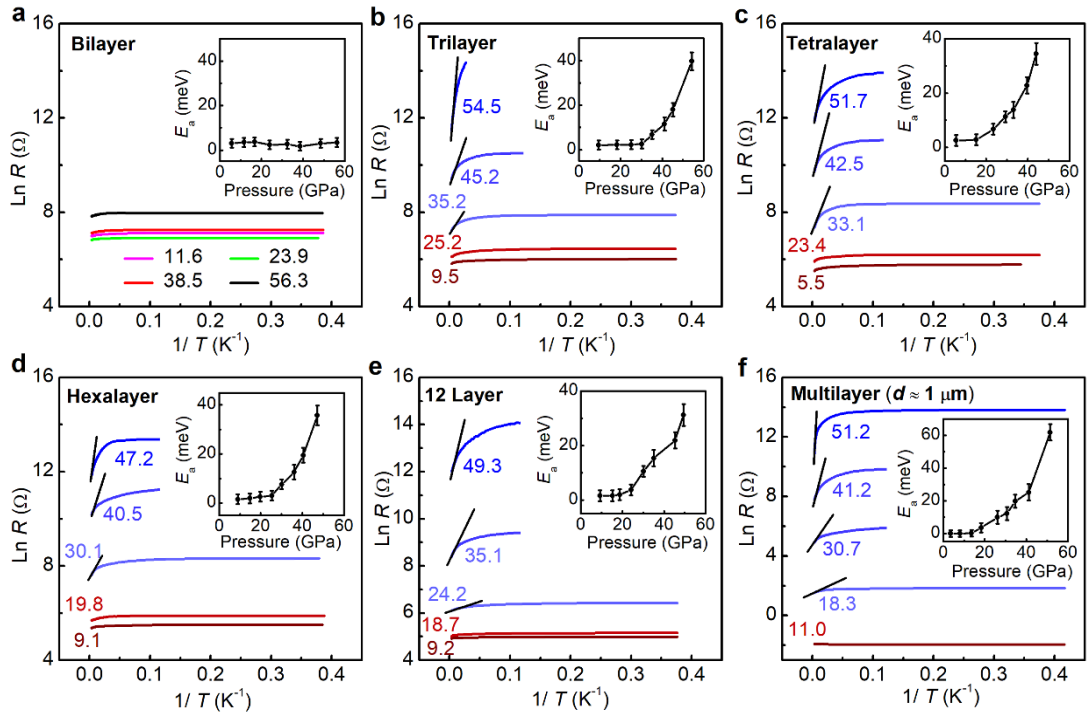

**Extended Data Figure 2 │ Temperature dependence of sheet resistance of bilayer, trilayer, tetralayer, hexalayer, 12-layer and multilayer graphene at representative pressures.** Arrhenius plots (Ln $R$ vs (1/$T$)) are adopted to fit the linear region at high temperature to obtain the activation energies, $E_a$ (inset). The activation energies of trilayer and thicker graphene increase dramatically after the graphene-diamane transition, in addition to the activation energies of bilayer graphene, which almost keeps its value of few meV up to 60.0 GPa (The highest pressure of the measurements).

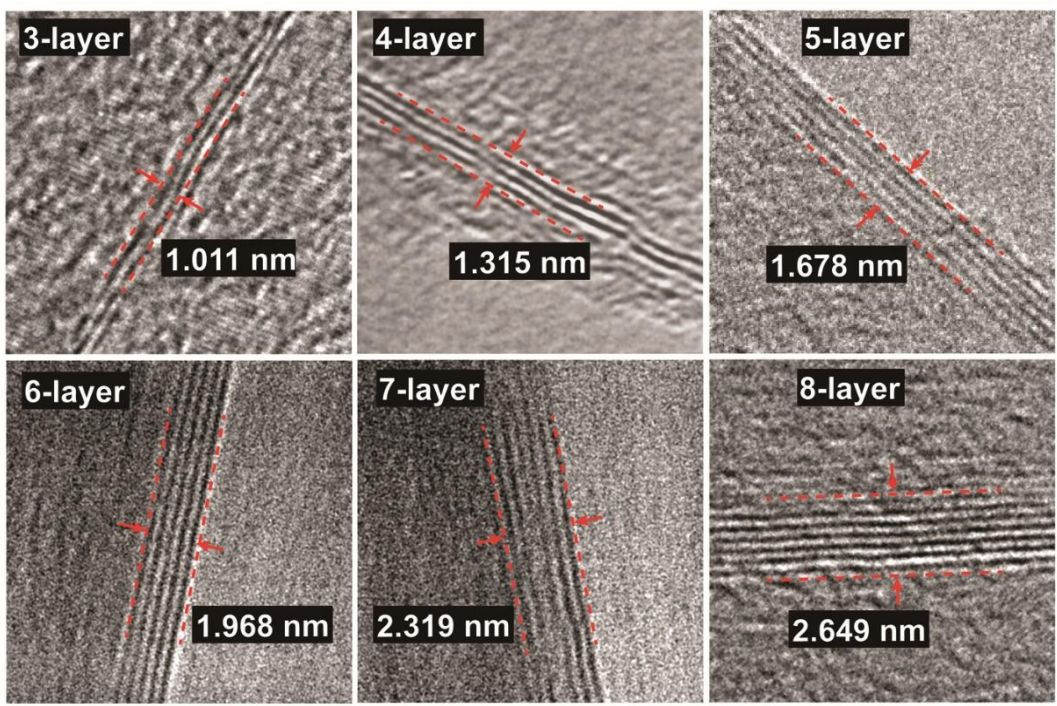

**Extended Data Figure 3 │ TEM images of mixed few-layer graphene viewing perpendicular to the c-axis.** The mixed few-layer graphene powders are provided by the Nanjing XFNANO Materials Tech Co., China. The average distance between each layer is about 3.34 Å, consistent with the thickness of monolayer graphene. The mixed powders are mostly made up of 3-8 layer graphene.

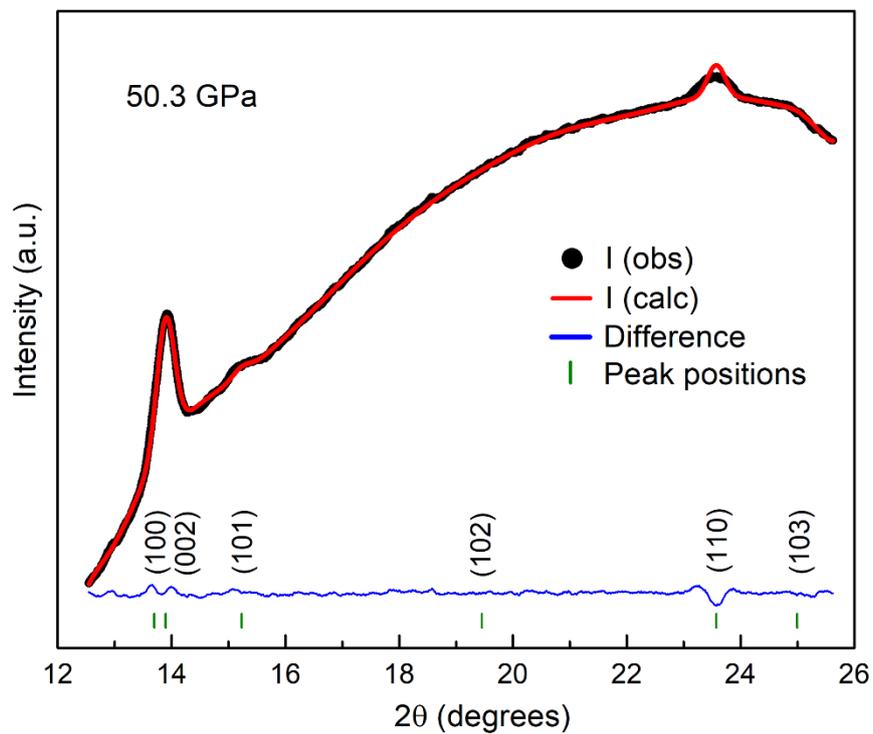

**Extended Data Figure 4 │ Le Bail fitting of the high-pressure phase with h-diamond structure.**

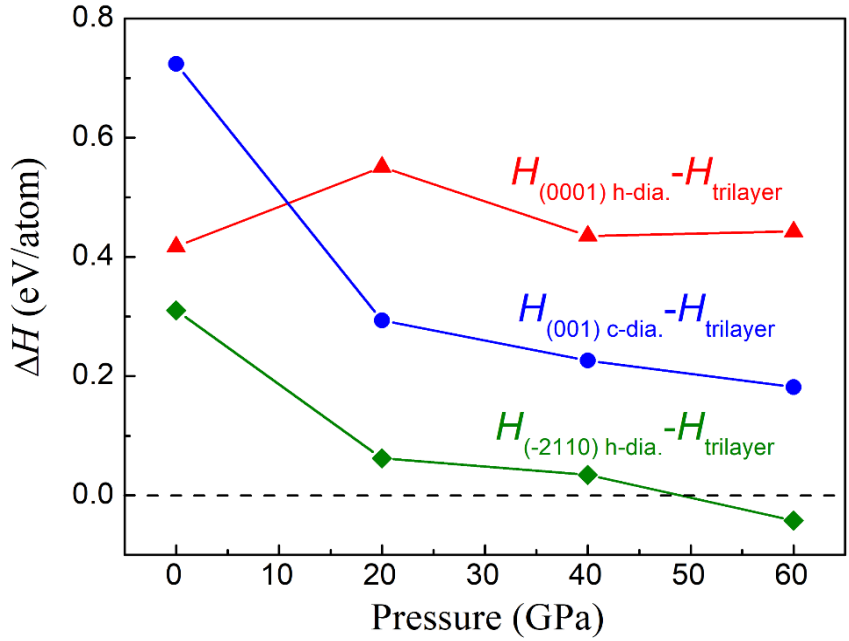

**Extended Data Figure 5 | Pressure dependence of enthalpies of h-diamane films on (-2110) and (0001) orientations, and cubic diamane film on (001) orientation transformed from trilayer graphene.** The structure searching results indicated that the (-2110) h-diamane film is more stable than the (0001) h-diamane film and (001) cubic diamane film and has a lower energy than that of trilayer graphene above 50 GPa. Both ABA and ABC stacked trilayer have been chosen as the starting structures. The calculations showed that trilayer with ABC stacking had a little higher energy than that with ABA stacking, and eventually transformed to an ABA stacking upon compression.

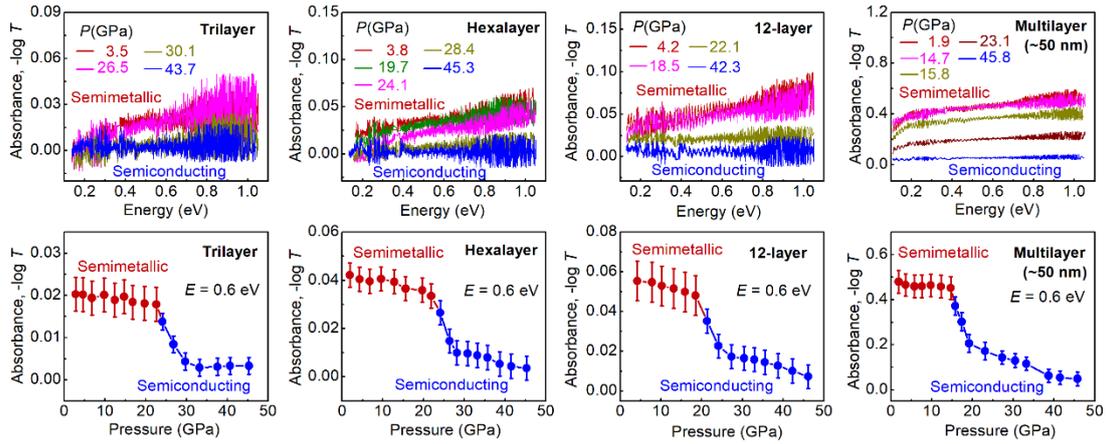

**Extended Data Figure 6 │ IR absorption of trilayer, hexalayer, 12-layer and multilayer graphene at pressures and the evolution of absorbance at photon energy of 0.6 eV under compression.** The absorbance of diamond and pressure medium is subtracted as background reference for each measurement. Sharp drops in absorbances are occurred followed the graphene to h-diamane film transition, clearly indicating that the opening of bandgap is larger than 1 eV.

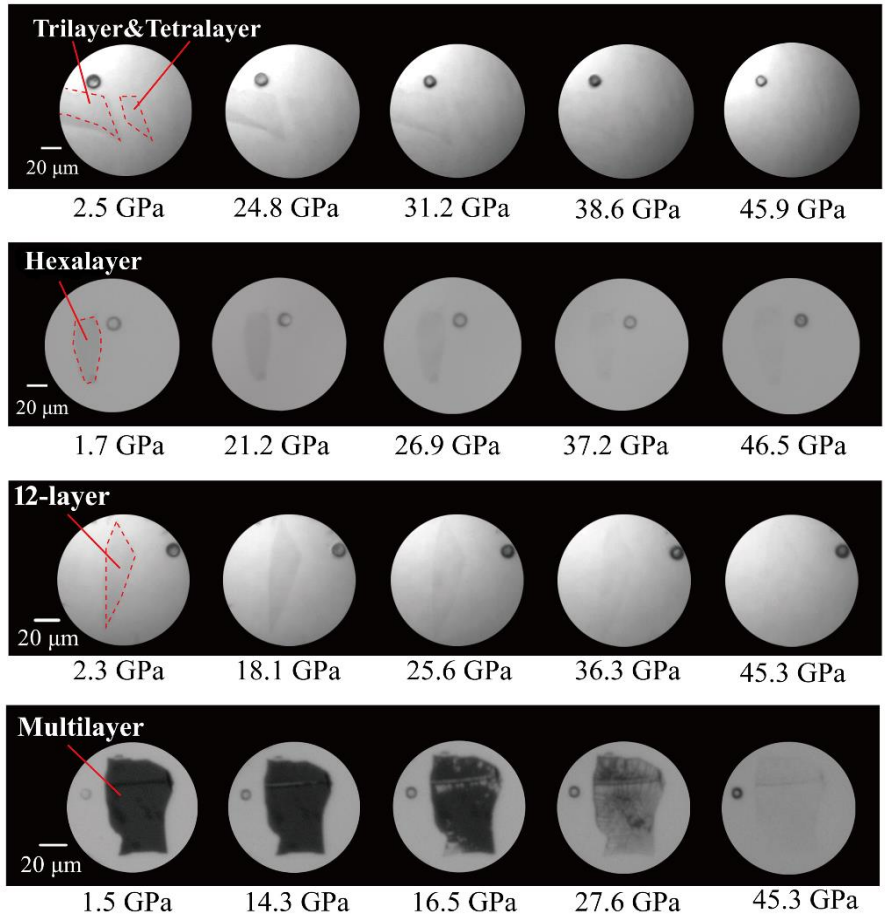

**Extended Data Figure 7 │ The optical microscope images of trilayer, tetralayer, hexalayer, 12-layer and multilayer graphene in transmission mode with white light source under compression.** The ruby ball (black dot) was used as the pressure indicator and visual reference. Daphne 7373 was loaded as the pressure medium. All the graphene samples become increasingly transparent above the transition pressure.

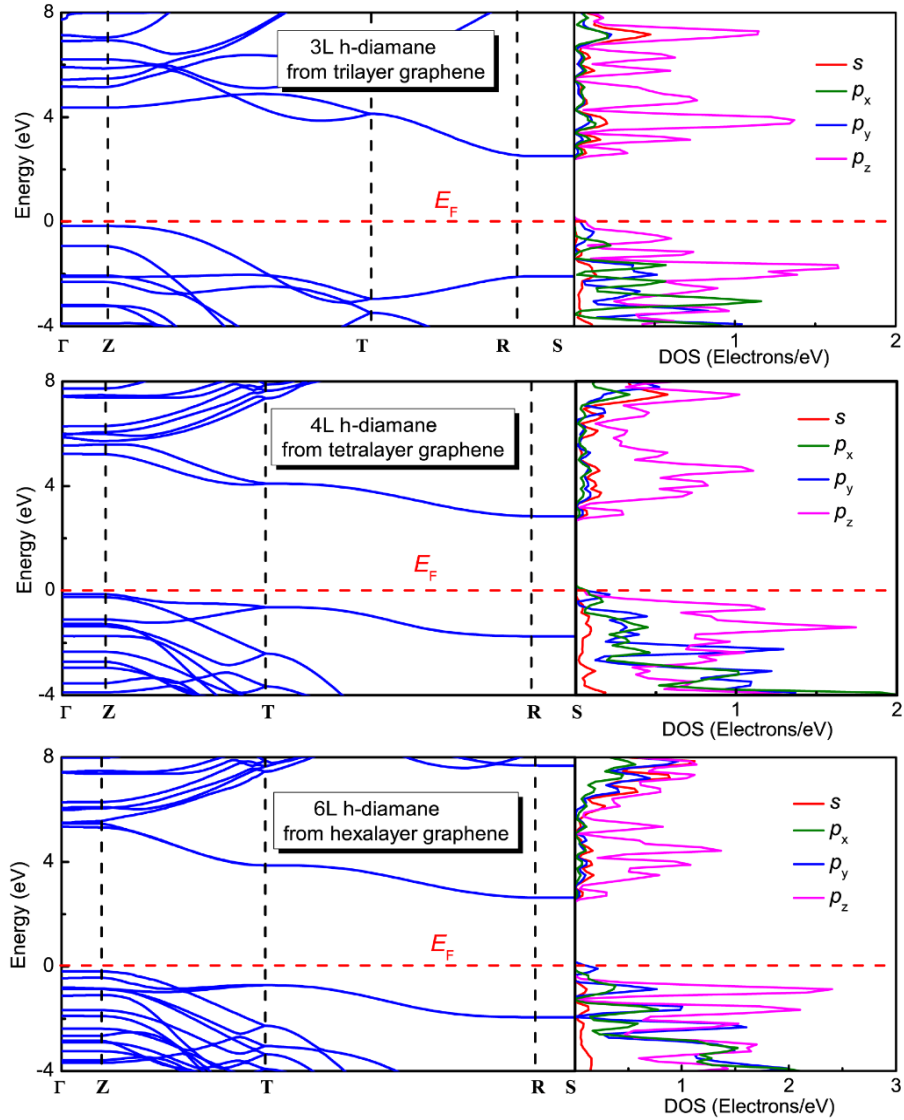

**Extended Data Figure 8 │ Band structure and Density of states (DOS) of diamane transformed from trilayer, tetralayer and hexalayer graphene.** The calculated value of bandgap is 3.0 ±0.1 eV, comparable with the optical adsorption results. The optical band gap of h-diamane looks not match the electrical activation energy obtained from resistance measurements, which is caused by the graphene/h-diamane intermediate phase. Above the transition pressure, such as 33.0 GPa for trilayer, the semimetallic trilayer graphene stats to transform to the semiconducting h-diamane but leaves a large percentage unchanged yet. The more conductive graphene dominates the

electrical properties. Hence, the resistance and activation energies show continuous increase rather than sharp jump although the bandgap is opened from 0 to 2.8±0.3 eV. With further compression, a large fraction of h-diamane formed, win out and dominate the electrical properties. Therefore, the resistance jumps out of the measurable range of instrument due to the large bandgap.